\documentclass[12pt,preprint]{aastex}

\usepackage{color}

\newcommand{\lSect}[1]{{\label{sec:#1}}}

\newcommand{\lEq}[1]{{\label{eq:#1}}}
\newcommand{\lTab}[1]{{\label{tab:#1}}}

\newcommand{\FIGFF}[2]{{\ref{fig:#2}{#1}}}

\newcommand{\Figure}[1]{{Figure~\FIGFF{}{#1}}}

\newcommand{\Sectff}[1]{{\ref{sec:#1}}}
\newcommand{\Sect}[1]{{\S~\Sectff{#1}}}

\newcommand{\Section}[1]{{\textsection \Sectff{#1}}}

\newcommand{\Table}[1]{{Table~\ref{tab:#1}}}

\newcommand{\powersep}{{\ensuremath{\times}}}

\newcommand{\pp}{\ensuremath{\,\%}}
\newcommand{\Msun}{\ensuremath{\,\mathrm{M}_\odot}}

\newcommand{\Zsun}{\ensuremath{\mathrm{Z}_\odot}}

\newcommand{\mue}{\ensuremath{\mu_{\mathrm{e}}}}
\newcommand{\HP}{\ensuremath{H_{\!P}}}
\newcommand{\Dz}{\ensuremath{D_{\!0}}}

\newcommand{\vbase}{\ensuremath{v_{\mathrm{base}}}}
\newcommand{\DOV}{\ensuremath{D_{\mathrm{OV}}}}
\newcommand{\enu}{\ensuremath{\epsilon_{\nu}}}
\newcommand{\enuc}{\ensuremath{\epsilon_{\mathrm{nuc}}}}

\newcommand{\Ep}[1]{{\ensuremath{10^{#1}}}}

\newcommand{\E}[1]{{\ensuremath{\powersep\Ep{#1}}}}

\newcommand{\isofont}[1]{{\mathrm{#1}}}
\newcommand{\isomass}[1]{{\ensuremath{\isofont{^{#1}}}}}
\newcommand{\isocharge}[1]{{\ensuremath{\isofont{_{#1}}}}}
\newcommand{\isotope}[3]{{\ensuremath{\isocharge{#1}\isomass{#2}\isofont{#3}}}}

\newcommand{\I}[2]{{\isotope{}{#1}{#2}}}
\newcommand{\El}[1]{{\I{}{#1}}}

\newcommand{\D}{\mathrm{d}}

\newcommand{\MCh}{{\ensuremath{\mathrm{M}_{\mathrm{Ch}}}}}
\newcommand{\Mcr}{{\ensuremath{\mathrm{M}_{\mathrm{crit}}}}}

\newcommand{\MESA}{\texttt{MESA}}
\newcommand{\MESAs}{\texttt{MESA star}}
\newcommand{\MminSN}{{\ensuremath{M^\mathrm{up ^\prime}}}}
\newcommand{\MminEC}{{\ensuremath{M^\mathrm{up}}}}
\newcommand{\MmaxSN}{{\ensuremath{M_\mathrm{max}}}}
\newcommand{\NSN}{{\ensuremath{N_\mathrm{SN}}}}

\newcommand{\Mcore}{{\ensuremath{M_\mathrm{core}}}}
\newcommand{\Zcr}{{\ensuremath{Z_\mathrm{cr}}}}

\newcommand{\X}[1]{#1}

\shorttitle{Metallicity Dependence of Type II SNe}
\shortauthors{Ibeling \& Heger}

\begin{document}


\title{
  The Metallicity Dependence of the Minimum Mass for Core-Collapse Supernovae}

\author{Duligur Ibeling}
\affil{Harvard College, Cambridge, MA 02138, U.S.A.}
\affil{Minnesota Institute for Astrophysics, 
School of Physics \& Astronomy, \\ University of Minnesota, 
Minneapolis, MN 55455, U.S.A.}
\email{duligur@gmail.com}
\and
\author{Alexander Heger\altaffilmark{1,2}}
\affil{Monash Centre for Astrophysics, 
School of Mathematical Sciences,\\ 
Building 28, M401, 
Monash University, Vic 3800, Australia}
\affil{Minnesota Institute for Astrophysics, 
School of Physics \& Astronomy, \\ University of Minnesota, 
Minneapolis, MN 55455, U.S.A.}
\altaffiltext{1}{Joint Institute for Nuclear Astrophysics}
\altaffiltext{2}{ARC Future Fellow}
\email{alexander.heger@monash.edu}





\begin{abstract}
  Understanding the progenitors of core collapse supernovae and their
  population statistics is a key ingredient for many current studies
  in astronomy but as yet this remains elusive. Using the {\MESA}
  stellar evolution code we study the dependence of the lower mass
  limit for making core collapse supernovae (SNe) as function of
  initial stellar metallicity.  We find that this mass limit is
  smallest at $[Z]\approx-2$ with a value of $\sim8.3\Msun$.  At
  $[Z]=0$ the limit is $\sim9.5\Msun$ and continues to rise with
  higher metallicity.  As a consequence, for a fixed initial mass
  function the supernova rate may be $20\pp$ to $25\pp$ higher at
  $[Z]=-2$ than at $[Z]=0$.  This affects the association of observed
  SN rates as a probe for the cosmological star formation rate, rate
  predictions for supernova surveys, and population synthesis studies.
\end{abstract}


\section{Introduction}

Most massive stars end their lives as core collapse supernovae (CCSN;
\citealt{c66}).  There is a lower mass limit below which a single star
of a given metallicity, $Z$, will not undergo such a supernova (SN;
\citealt{e04,h03,s09}). Below this limit, a star ejects its envelope
as an asymptotic giant branch (AGB) star, forms a planetary nebula,
and becomes a white dwarf \citep{e04}; the remaining mass of the star
will be less than the Chandrasekhar mass, $\MCh\approx1.46\Msun$.
Above the limit, however, the star builds up an iron core by nuclear
fusion until the mass of the core exceeds a critical mass, $\Mcr$
\citep{e04, h03}.  In such a star core collapse occurs: the iron core
will fall inwards until repulsive nuclear forces stop its collapse.

The actual value of this lower mass limit depends greatly on a variety
of factors \citep{n88}, including the numerical simulations code used
and how the relevant physics is modeled, e.g., the mass loss rates,
mixing processes, convective boundary layers and semiconvection, which
change the resulting interior composition structure of the star.
Rotation may also affect the limit \citep{e04}, but be we do not
explore this in the present study.  Metallicity-independent values of
$8\Msun-10\Msun$ are often used in the literature \citep{h03}, though
we expect that it should be affected by metallicity \citep{c93,e04}.
Current models \citep{e04} of the lower mass limit as a function of
$Z$ have been made with only a limited resolution in $Z$ and cover a
rather limited $Z$ regime.  Observational determinations are too
uncertain to explore the lower limit precisely. \cite{s09} determines
from a combination of supernovae detections and the Salpeter initial
mass function (IMF) that the lower limit lies between $7\Msun$ and
$9\Msun$, whereas \cite{c93} focused on the dependence over a wide
range of $Z$ values.  These investigations, however, have not provided
detailed information on how the lower limit depends on $Z$.

A star with initial mass lower than that required for a classical CCSN
may also undergo core collapse triggered by electron captures
\citep{Nom84}; this is expected to occur for a narrow range of initial
stellar masses \citep{m80, m87, p08}. \cite{p08} estimates this range
to be between $0.25\Msun$ and $0.65\Msun$.  We, however, do not
explore this case; we instead focus on the minimum mass required for
classical core collapse events.  We also do not consider binary stars
in which the mass limits will be altered by interaction including mass
transfer and accretion.  The minimum mass required for a star to still
be able to ignite carbon burning and possibly become an electron
capture SN is denoted $\MminEC$, while that required for a star to
undergo a classical core collapse event not triggered by electron
captures with an ONe degenerate core is denoted $\MminSN$.  Thus, we
focus on $\MminSN(Z)$.

We present results from a grid of non-rotating stellar models that map
the minimum supernova mass as a function of metallicity.
In \Section{simulations} we describe the models used and the physics
relevant to supernova progenitors.  Then in \Section{results} we give
the results from the grid of stellar models and provide a
metallicity-dependent fitting function for the minimum mass for
supernovae, $\MminSN(Z)$.  We discuss the implications of the results
on cosmology and galactic chemical evolution due to changes in the
number of SNe with $\MminSN(Z)$ in
\Section{discussion}.  The discussion includes a brief summary of
uncertainties and physical effects that cause the observed trends.
Our conclusions are given in \Section{conclude}.

\section{Setup and Simulations\lSect{simulations}}

We used {\MESA} (Modules for Experiments in Stellar Astrophysics,
\texttt{http://mesa.sourceforge.net/}) code revision 3290.  {\MESA} is
a modern, open source package for computational astrophysics
\citep{p11} and includes a 1D stellar evolution code, {\MESAs} whose
parts have been tested internally and verified with well-known
evolution results.

We computed a grid of non-rotating stellar models, with composition
scaled linearly according to $Y=0.24+2Z$.  For the relative mass
fractions of metals we use \citet{g98} and \cite{a89} for \El{Li},
\El{Be}, and \El{B}.  The computation grid covers $24$ values of $Z$
between $0$ and $0.04$, and the initial masses ranged from
$M=8.2\Msun$ to $M=10\Msun$.  For each model we follow the evolution
from the pre-main sequence to the point where we can determine whether
a core collapse supernova should result.

Whereas recent solar abundance determinations \citep{asp09,LPG09} have
different abundance ratios and, foremost, a different absolute
metallicity, this should have little effect on our overall
conclusions.  Additionally, the use of scaled solar metallicity is a
common simplification, and whereas it does not account for details of
the galacto-chemical evolution of the different species, nor the
spread of abundance ratios for a given metallicity for different
environments or times, it should still approximate the key properties
of how varying metallicity changes the supernova mass limit.

\subsection{Mass loss} 
\lSect{massloss}

Mass loss can strongly affect stellar evolution and the final fate of
the star, especially for high mass and high metallicity (e.g.,
\citealt{h03}).  Here we applied the mass loss rates of \citet{d88} to
the entire range of stars.  Despite their age they are still
considered adequate \citep{e04}.  These rates are empirical but they
agree well with the theoretical rates of \citeauthor{v00}
(\citeyear{v00}, \citeyear{v01}; \citealt{e04}).  The mass loss rates
were scaled by a constant efficiency factor $\eta = 0.8$ to adjust the
empirical rates such that evolutionary tracks for non-rotating models
match observational data \citep{m01}.

We scale the mass loss rates with metallicity using
$\dot{M}\left(Z_0\right)=\dot{M}\left(\Zsun\right)\sqrt{Z_0/\Zsun} $
\citep{k00} where we use the same $\Zsun=0.019$ \citep{a89} as was
used in the underlying mass loss rate of \citet{d88}, and $Z_0$ is the
initial metallicity of the star.  We added this scaling to {\MESAs}
for our study.

\subsection{Overshooting} 

Convective overshooting \X{increases} the size of
the helium core \citep{s97} and thereby affects the mass limit for
supernovae.  We use the exponential diffusive model based on the
prescription of \cite{f96}.  The diffusive coefficient, $\DOV$, is
given by $\DOV=\Dz\,\exp\left\{-2z/\left(f\HP\right)\right\}$, where
$z$, $\HP$, and $\Dz$ are determined by the particular overshooting
region: $z$ is the location from its boundary, $\HP$ is the
pressure scale height at the convective boundary, and
$\Dz=\vbase^{2}\,t$.  Here $\vbase$ is a typical velocity at the
boundary of the convective \X{zone}, $t=\HP/\vbase$ a typical
timescale \citep{f96}, and $f$ is a free parameter that governs the
efficiency of the overshoot mixing.
\X{For all models we adopt a value of $f = 0.016$, following \cite{h00}.}

\subsection{Determining the fate of the star}

The evolution past central helium burning of a star with initial mass
between $8\Msun$ and $10\Msun$ generally proceeds along the AGB.
This begins by dredging up the helium shell above the core before
usual helium shell flashes set in \citep{h05, i83}.  A star that
produces a CCSN, however, will be halted in this evolution usually
before the shell flashes by igniting neon burning.  Once neon has
ignited, the core typically burns all the way to iron and a core
collapse ensues.

The maximum mass of a stable white dwarf is
$\MCh=5.83\,\mue^{-2}\Msun$ \citep{c35}.  If the core consists
entirely of $\I{12}C$ and $\I{16}O$, then $\mu_{\mathrm{e}}=2$ and
$\MCh=1.46\Msun$.  But if the core is composed of $\I{56}{Fe}$, then
$\mue=2.15$ and $\MCh=1.26\Msun$.  Hence it is necessary to consider
the composition of the core to obtain a useful value for $\MCh$.  For
our models we have found that a value $\MCh=1.38\Msun$ was a good
indicator for igniting advanced burning phases, and this value is
consistent with those used by other studies \citep{e04}.  It
corresponds to a $\mue\approx2.055$.

\citet{e04} find that the final fate of the star depends greatly on
the second dredge-up, since the second dredge-up directly impacts the
mass and chemical composition of the core by removing material of a
certain composition from the core. The interplay of dredge-up and
burning processes determines whether the cores can reach $\MCh$ and
hence the final fate of the stars. For example, for stars of about
$7\Msun-9\Msun$ initial mass, dredge-up reduces the core size below
$\MCh$ so that a SN will not result; for stars of initial mass
$\gtrsim11\Msun$ dredge-up will not occur, and nuclear burning
increases the core mass until it cannot support itself and a SN
results \citep{e04}.

Therefore simulations were run at least until the second dredge-up was
finished by the time we determined the final fate.  We assumed the
outcome will be a white dwarf (WD) when: \emph{i)} $\enu>\enuc$ in the
core, i.e., energy loss due to neutrino emissions is greater than
energy generation by nuclear burning, so that the core continues to
cool; and \emph{ii)} $\Mcore<1.38\Msun$.

These two criteria imply that further burning in the core will not
ignite.  Here we do not consider the case of electron capture
supernovae (ECSN) in super-AGB stars \citep{p08}.  In that case, an
ONe core collapses due to electron captures on $^{20}\mathrm{Ne}$ and
$^{24}\mathrm{Mg}$ before onset of \El{Ne} burning \citep{m80, m87}.
Those would allow supernovae from stars with initial masses below the
lower limit for core collapse supernovae as studied here.

Conversely, we use the following criteria as indicators that a star
becomes a core collapse SN: \emph{i)} The onset of \El{Si} burning,
the final stage before core collapse; and \emph{ii)} $\Mcore >
1.38\Msun$.

It may not be necessary to wait for \El{Si} burning, since after the
onset \El{Ne} burning the core should proceed to \El{Si} burning.  To
assure this, our models were run to \El{Si} burning, nevertheless.
Since the \El{Si} burning timescale ranges from about $18\,$d
\citep{w05} to about $1\,$yr and since this is much shorter than the
dredge-up timescales, the core mass will not shrink further.  The
second criterion ensures that core is large enough to undergo core
collapse.  We require that the stellar core must be well established
and therefore the models must be run at least until they finish the
second dredge-up.

We stopped the evolution simulation when both criteria for either one
of the stellar fates (WD or SN Type II) were met and allowed us to
determine the fate of the model.

\section{Results\lSect{results}}

\begin{deluxetable}{ll|ll|ll}
\tablecaption{Summary of Model Fates\lTab{models}}
\tablewidth{0pt}
\tablecolumns{4}
\tablehead{
\colhead{$Z$} &
\colhead{$\MminSN$} &
\colhead{$Z$} &
\colhead{$\MminSN$} &
\colhead{$Z$} &
\colhead{$\MminSN$}
\\
\colhead{} &
\colhead{(\Msun)} &
\colhead{} &
\colhead{(\Msun)} &
\colhead{} &
\colhead{(\Msun)}
}
\startdata
0        & 9.2  & 1\E{-7}  & 8.8  & 1\E{-3}  & 8.45 \\
1\E{-11} & 9.25 & 1\E{-6}  & 8.7  & 2.5\E{-3}& 8.65 \\
1\E{-10} & 9.25 & 4\E{-6}  & 8.56 & 5\E{-3}  & 8.95 \\
3\E{-10} & 9.35 & 7\E{-6}  & 8.55 & 7.5\E{-3}& 9.05 \\
5\E{-10} & 9.5  & 1\E{-5}  & 8.47 & 1.5\E{-2}& 9.35 \\
1\E{-9}  & 9.5  & 2\E{-5}  & 8.43 & 1.9\E{-2}& 9.5  \\
3\E{-9}  & 9.3  & 5\E{-5}  & 8.47 & 3\E{-2}  & 9.6  \\
1\E{-8}  & 9.05 & 1\E{-4}  & 8.35 & 4\E{-2}  & 9.85   
\enddata
\end{deluxetable}

In \Table{models}, we list for each metallicity, $Z$, the minimum mass
$\MminSN$ of a model that produced an SN Type II.  A
metallicity-dependent transition mass, $\MminSN(Z)$, can be fitted by
\begin{equation}
   \frac{\MminSN(Z)}{\Msun} = \left\{
     \begin{array}{lr}
       \sum_{i=0}^6a_i[Z]^i&:[Z]\ge-8.3\\
\\
       9.19&:[Z]<-8.3
     \end{array}
   \right\}\pm0.15
\lEq{fit}
\end{equation} 
with coefficients $a_i$ given by
\begin{equation}
  \begin{array}{lr}
    a_0=+9.40858\\
    a_1=+1.14548\\
    a_2=+3.96411\E{-1}\\
    a_3=+2.96185\E{-2}
  \end{array}
  \begin{array}{lr}
    a_4=-8.79237\E{-3}\\
    a_5=-1.96134\E{-3}\\
    a_6=-1.11999\E{-4}\;.
  \end{array}
\lEq{coeff}
\end{equation}
For $[Z]\ge-8.3$, where $[Z]=\log{Z/\Zsun}$, this formula has a
$1\,\sigma$ deviation of only $0.056\Msun$ ($0.112$ for the entire
range).

\begin{figure}
\begin{center}
\plotone{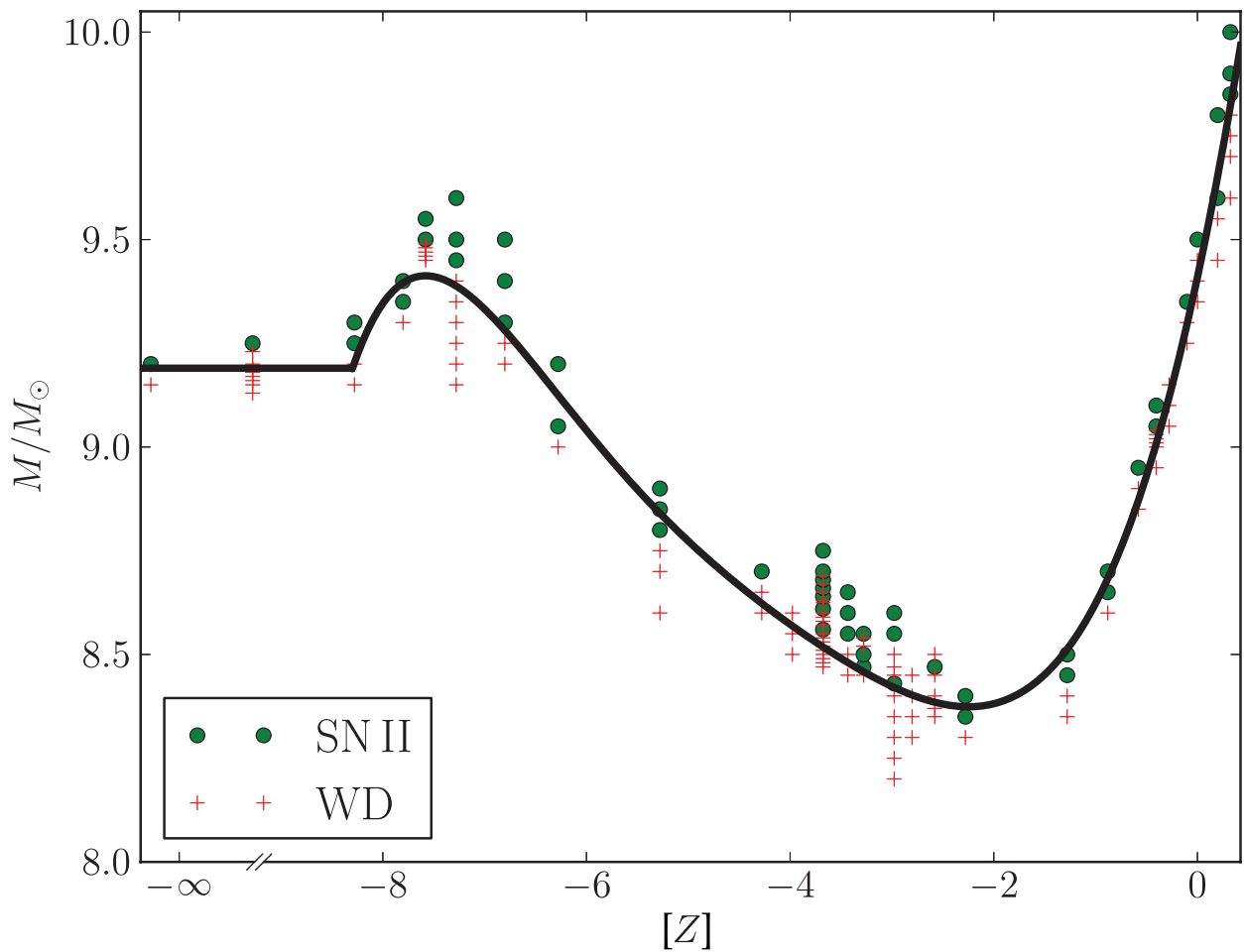}
\caption{Fate of stars as a function of mass and
    metallicity. \textit{Filled green circles} indicate models that
    make core collapse supernovae, \textit{red crosses} show models
    that do not, and the \textit{black curve} indicates our fit.
 \label{fig:ZM}}
\end{center}
\end{figure}

We find that for $[Z]<-8.3$ the lower mass limit becomes metallicity-independent and is best fit by a constant value of $9.19\Msun$.
\Figure{ZM} shows the fate for each model we computed as a function of
its initial $M$ and $Z$ and our fitting function $\MminSN(Z)$.

\section{Discussion\lSect{discussion}}

Our results are consistent with both observations \citep{s09} and the
current paradigm \citep{e04, h03} that the minimum supernova mass lies
between $8\Msun$ and $10\Msun$.  Specifically, previous findings
\citep{e04, n88} that the fate of the star depends on the evolution
during the second dredge-up are confirmed.  For all models, an ONe
core grew during dredge-ups.  We could determinate whether the star
makes a supernova after the second dredge-up.  \citet{e04} note that
this behavior is because of early \El{C} burning, which is confirmed
by our models, including off-center ignition of \El{C} burning.  In a
few cases the more advanced burning stages (usually Si burning) did
not occur.  In those cases we assumed that the star ejects its
envelope and peacefully becomes a WD.  Once \El{Si} burning ignites,
however, the star will evolve to core collapse.

\subsection{Uncertainties}

\X{We do not find a sharp transition mass in our simulations.
  Instead, for a given $Z$, we find a switching back and forth between
  making SNe and WDs several times as the initial mass is increased.
  Most of this might be attributed to numerical noise.  We note,
  however, that the non-linear nature of stellar structure, e.g., the
  interaction of different shell burning phases during carbon burning,
  may cause similar phenomena.  But we cannot quantify the relative
  magnitude of this effect, if present.  It is included in our error
  estimate.}  In ``real'' stars, additionally, convection will be
chaotic–--there is ``weather'' inside stars–--but this is an effect
beyond the realm of our simulations, next to rotation, binary stars,
etc.  Based on our detailed studies, e.g., for $[Z]=-3.7$, we estimate
an uncertainty of the transition mass of about $\pm0.15\Msun$ for all
values of $Z$, \X{excluding systematic errors due to the model}.

\subsection{Impact of Metallicity on Evolution}

At the highest metallicities we see the effect of increasing mass loss
(\Sect{massloss}) that reduces the mass of the star and hence shifts
the mass limit upward.  Additionally, an increase of opacity due to
higher metal contact can also lead to more efficient dredge-ups that
reduce the size of the core.  The increase of the transition mass at
low metallicities ($[Z]\lesssim-2$), where mass loss becomes
unimportant, is dominated by a decrease in opacity.  Therefore a
smaller fraction of the core will be convective, leading to larger
transition mass as $[Z]$ decreases below $-2$.  Below $[Z]\lesssim-8$
the initial abundance of CNO isotopes is so low that it is
insufficient to efficiently drive hydrogen burning: the star contracts
until carbon is made by fusing helium in a primary way, then
re-expands, hence the evolution becomes $Z$-independent for even lower
$Z$.

Whereas $\MminSN(Z)$ does depend on the specific stellar model
code, the choice of physics used, and numerical implementation, as is
well known, we expect that the general shape of the function arises
from the underlying stellar physics and should not depend on the
specific simulations used.

\subsection{Impact on supernova rates}

Using the \citet{s55} IMF we can compute the number of stars that make
SNe, 
$$
\NSN=\xi_0\int_\MminSN^\MmaxSN\,M^{\,-2.35}
\D M\,,
$$ 
where $\xi_0$ is a normalization constant.  The lower limit,
$\MminSN$, of the integral was explored here as a function of $Z$.  A
conservative upper limit for the maximum initial mass of a supernova
progenitor, $\MmaxSN$, may be as high as $100\Msun$.  Stars of higher
mass are very rare.  More realistic is a typical transition mass from
SNe to collapse without SN display at $20\Msun-30\Msun$
\citep{h03,o11,u12} to as low as $16\Msun$ based on recent
observational studies \citep{s09}. Therefore a rough estimate of a
typical $\MmaxSN \approx 20\Msun$ is not unreasonable, and we will use
this as our reference case.  Similar to $\MminSN$, there may not be a
unique transition mass \citep{h03,o11,u12}.  Whereas the upper limit
may depend on metallicity \citep[e.g.,][]{ZWH08}, this dependence is
very uncertain and beyond the purpose of this paper.  \citet{w02},
Fig.~4, show that between $16\Msun$ and $27\Msun$ there is little
difference in core mass, and hence anticipated SN properties, for
stars of $[Z]=-4$ and $[Z]=0$ and the same initial mass.  Therefore we
will consider here only a constant upper mass limit.

Although the IMF itself could depend on metallicity as well, there is
no overwhelming observational indication that it does \citep{k02}.
From theoretical considerations it was suspected that below a critical
metallicity $\Zcr\sim\Ep{-6}$ only very massive ($30\Msun-300\Msun$)
stars formed \citep{c05} that would not form supernovae and collapse
to black holes instead.  Only above $\Zcr$ would a ``normal'' IMF set
in.  The recently reported very metal-poor ($Z\le7\E{-7}$) and
low-mass star \citep{c11} however calls into question the validity of
$\Zcr$.  Recent work on nucleosynthesis from Pop III stars
\citep{HW10} also does not provide strong support for a change in the
IMF, at least not for massive stars.  Hence we assume a
metallicity-independent IMF.

The $Z$-dependence of $\MminSN$ leads to a large change in $\NSN$ due
to the high weight of the IMF at low $M$.  Using our preferred value
of $\MmaxSN=20\Msun$, our value $\MminSN=8.35\Msun$ at $[Z]=-2.3$
(\Table{models}) this leads to a $25\pp$ increase of $\NSN([Z]=-2.3)$
compared to $\NSN([Z]=0)$ with $\MminSN(\Zsun)=9.2\Msun$
(\Table{models}).  Assuming the very conservative value of
$\MmaxSN=100\Msun$ for comparison, the change in $\NSN$ still is
$20\pp$.  The magnitude of the enhancement is not very sensitive to
the upper mass limit as long as it is $\gtrsim20\Msun$.  The
uncertainty in $\MmaxSN$ dominates over the impact of the numerical
noise in $\MminSN$ on the estimate of $\NSN$.

More important for cosmological applications is to know $\NSN$ as a
function of redshift \citep{LF09}.  This would allow us to find
the number of supernovae formed at each epoch of the history of the
universe.  To make this connection, we need to know the distribution
of $Z$ and star formation rate as a function of redshift.  These
parameters are all known to be connected to some extent; for example,
\cite{m10} give a fundamental metallicity relation (FMR) that connects
star formation rate, metallicity, and galaxy mass, but no better than
$\pm12\pp$.  A detailed assessment of this, however, is beyond the
scope of this work.  The effect we present here is in contrast to the
recently found lack of core collapse SNe at high redshift \citep{hor11}.

Similar to the SN rates, the variation of the lower mass limit with
metallicity changes the ratio of neutron star (NS) to black hole
(BH) formation used in population synthesis studies \citep{fry99}.
The effective nucleosynthesis yields from supernova used in
galacto-chemical evolution models will be changed by allowing a larger
contribution from low-metallicity stars.

\vspace{-0.5\baselineskip}

\section{Conclusions\lSect{conclude}}

We have computed for the first time the minimum mass for stars to make
classical core collapse supernovae, $\MminSN$, as a function of
metallicity, $Z$, at a high level of detail.  Our results generally
agree with previous estimates by \citet{e04, h03} but significantly
improve the quantitative results and detailed $Z$ dependence.

We find an increase of $\MminSN$ for $[Z]\gtrsim-2$, resulting in a
decrease in the fraction of stars making CCSNe for a fixed IMF.
Compared to a constant lower mass limit for supernovae this
constitutes an increase in the supernova rate by $20\pp$ to $25\pp$ at
low metallicity.  Only for $[Z]\lesssim-2$ does the SN rate increase
again.  This has significant impact on interpreting observed SN rates
as a probe of star formation history, on predictions of expected SN
rates for upcoming surveys, and for NS/BH ratios for population
synthesis studies.

Future work should include stellar rotation, binary stars, the mass
range required for ECSN from super-AGB stars, including how $\MminEC$
changes relative to $\MminSN$ as a function of metallicity, and a
better understanding of the upper mass limit for supernovae.  We need
to connect the results to galactic parameters in order to better
quantify the relative number of supernovae, neutron stars, and black
holes as a function of redshift.  Much of the underlying theory for
these, however, is still very uncertain at the present.

\section{Acknowledgments}

We thank Bill Paxton and the {\MESA} user group for their advice and
support.  This research was supported by NSF through grant
AST-1109394.  AH acknowledges support from the DOE Program for
Scientific Discovery through Advanced Computing (SciDAC;
DE-FC02-09ER41618), by the US Department of Energy under grant
DE-FG02-87ER40328, and by the Joint Institute for Nuclear Astrophysics
(JINA; NSF grant PHY02-16783), and an Australian ARC Future
Fellowship.



\end{document}